\documentclass[a4paper]{article}
\usepackage[a4paper, total={6in, 8in}]{geometry}
\usepackage{amssymb}
\usepackage[figuresright]{rotating}
\usepackage{authblk}

\begin{document}

\title{Time calibration with atmospheric muon tracks in the ANTARES neutrino telescope}

\date{}

\author[1]{S.~Adri\'an-Mart\'inez}
\author[2]{A.~Albert}
\author[3]{M.~Andr\'e}
\author[4]{G.~Anton}
\author[1]{M.~Ardid}
\author[5]{J.-J.~Aubert}
\author[6]{B.~Baret}
\author[7,b]{J.~Barrios-Mart\'{i}}
\author[8]{S.~Basa}
\author[5]{V.~Bertin}

% ...
\affil[b]{\scriptsize{Corresponding author}}
\affil[1]{\scriptsize{Institut d'Investigaci\'o per a la Gesti\'o Integrada de les Zones Costaneres (IGIC) - Universitat Polit\`ecnica de Val\`encia. C/  Paranimf 1 , 46730 Gandia, Spain.}}
\affil[2]{\scriptsize{GRPHE - Universit\'e de Haute Alsace - Institut universitaire de technologie de Colmar, 34 rue du Grillenbreit BP 50568 - 68008 Colmar, France}}
\affil[3]{\scriptsize{Technical University of Catalonia, Laboratory of Applied Bioacoustics, Rambla Exposici\'o,08800 Vilanova i la Geltr\'u,Barcelona, Spain}}
\affil[4]{\scriptsize{Friedrich-Alexander-Universit\"at Erlangen-N\"urnberg, Erlangen Centre for Astroparticle Physics, Erwin-Rommel-Str. 1, 91058 Erlangen, Germany}}
\affil[5]{\scriptsize{CPPM, Aix-Marseille Universit\'e, CNRS/IN2P3, CPPM UMR 7346, 13288, Marseille, France}}
\affil[6]{\scriptsize{APC, Universit\'e Paris Diderot, CNRS/IN2P3, CEA/IRFU, Observatoire de Paris, Sorbonne Paris Cit\'e, 75205 Paris, France}}
\affil[7]{\scriptsize{IFIC - Instituto de Física Corpuscular, Parque Científico, C/Catedrático José Beltrán 2, E-46980 Paterna, Spain}}
\affil[8]{\scriptsize{LAM - Laboratoire d'Astrophysique de Marseille, P\^ole de l'\'Etoile Site de Ch\^ateau-Gombert, rue Fr\'ed\'eric Joliot-Curie 38,  13388 Marseille Cedex 13, France}}
% ...

\maketitle

\begin{abstract}
The ANTARES experiment consists of an array of photomultipliers distributed along 12 lines and located deep underwater in the Mediterranean Sea. It searches for astrophysical neutrinos  collecting the Cherenkov light induced by the charged particles, mainly muons, produced in neutrino  interactions around the detector. Since at energies of $\sim$10 TeV the muon and the incident neutrino are almost collinear, it is possible to use the ANTARES detector as a neutrino telescope and identify a source of neutrinos in the sky starting from a precise reconstruction of the muon trajectory. To get this result, the arrival times of the Cherenkov photons  must be accurately measured. A to perform  time calibrations with the precision required to have  optimal performances of the instrument is described. The reconstructed tracks of the atmospheric muons in  the ANTARES detector are used  to determine the relative time offsets between photomultipliers. Currently, this method is used to obtain the time calibration constants for photomultipliers on different lines at a precision level of 0.5 ns. It has also been validated for calibrating photomultipliers on the same line, using a system of LEDs and laser light devices.
\end{abstract}


\section{Introduction}
\label{sec:intro}
The ANTARES neutrino telescope \cite{bib:antares_nim}  aims at the exploration of the high-energy Universe by using neutrinos as cosmic probes. One of the main goals of the experiment is the discovery of point-like sources of cosmic neutrinos. Typically, searches for sources of neutrinos are performed using the reconstructed directions of selected events. 
In this sense, a good angular resolution, i.e., the angle between the direction of the reconstructed muon and of the neutrino, is of great importance to better discriminate between the signal and the background. At energies $\sim10$ TeV the median value of the angular resolution in the ANTARES detector, as  determined with simulations, is $\sim 0.4^{\circ}$ \cite{bib:antares_PS}. 

In the energy range of interest, the angular resolution is driven by the reconstruction accuracy. Since the reconstruction algorithm depends on the precise measurement of the photon arrival 
times on the photomultipliers (PMTs), an accurate detector time calibration is crucial to guarantee the best performance of the telescope. 
In particular, the precision in the measurement of the relative PMT timings is required to be $\sim1$ ns \cite{bib:antares_tcalib}.

In this paper, the  methods to perform the time calibration of the detector at this level of accuracy and to determine the calibration constants used in the physics analyses of the ANTARES collaboration are presented. The method is based on the reconstructed trajectories of the down going atmospheric muons, which are the main source of physical triggers for a neutrino telescope. This calibration method does not require the physics data acquisition to be stopped and  does not rely on additional electronic devices. It was initially implemented for the first ANTARES point-source analysis \cite{bib:first_point}.

The paper is structured as follows. In section 2 a brief description of the ANTARES detector is presented. The calibration method  is introduced in Section 3. In the fourth Section the reconstruction method used in the point-like source search, and in the majority of the physics analyses in ANTARES, is described.  In Sections 5 and 6 the derivation of the calibration constants is described in detail. In Section 7 the results are discussed and compared to the results obtained with an independent system of calibration that uses an array of optical beacons \cite{bib:antares_beacons}. A summary and the conclusions are given in Section 8.

\section{The ANTARES detector}
ANTARES is the first deep-sea neutrino telescope in operation in the world. Anchored at a depth of 2475 m in the Mediterranean Sea, 40 km off the coast of Toulon (France), 
the detector  comprises 885 PMTs distributed on a three-dimensional array made up by 12 flexible lines, each of them 480 m long, which are arranged on an octogonal layout with an interline separation of 60-75 m. Each line holds triplets of PMTs and the control electronics boards needed for the power supply and data control at this level, at 25 vertical positions (storeys). The first storey of each line is placed 100 m above the seabed and the distance between adjacent storeys is 14.5 m. Each PMT (10'' photocathode, model R7081-20 of Hamamatsu \cite{bib:antares_pmt}), hosted in a high-pressure-resistant glass sphere together with high-voltage power supplies and internal calibration tools, constitues an optical module (OM) \cite{bib:antares_om}. 

The electronics boards are enclosed in a titanium container, the local control module. The main  electronic components are the analogue ring sampler (ARS) circuits  \cite{bib:antares_ars}, which perform the digitisation of the electronic signals produced in the PMTs, providing information on amplitude, arrival time and signal shape. Two ARSs per OM work in a token ring configuration to minimise the acquisition dead time. Inside each ARS, an amplitude-to-voltage converter is used for the signal charge integration, and a time-to-voltage converter  allows the measurement of hit times with a sub-nanosecond precision. 

The 12 ANTARES detection lines are connected to a junction box that, 
through an electro-optical cable about 40 km long, links the detector to 
the shore station and provides the power and control signals needed for the full apparatus. 
All the signals passing a threshold condition, referred to as L0 trigger and typically set to 0.3 photoelectrons, activate the digitisation and the sending of the hit time and the integrated charge of the signal to the shore station \cite{bib:antares_daq}. They are then processed with a cluster of computers by applying a data-filter algorithm looking for space-time correlated hits consistent with a specific physics 
signal  to mitigate the effect of optical background.
Finally, the detector includes a set of additional instruments for calibration purposes, such as hydrophones 
for the positioning system and optical beacons for timing calibration. In particular, the 
optical beacons system consists of a series of LED and laser light devices which are distributed 
along the detector to illuminate the PMTs. 

\section{Time calibration in ANTARES}
\label{sec:time_calib}
An accurate muon track reconstruction requires a precise determination of the arrival times of the photons on the PMTs. In particular, all the PMTs have to be synchronised to within $\sim$1 ns. The main uncertainties on the measured hit times come from the transit time spread of the PMTs and the optical properties of sea water.
In the ANTARES experiment, the first time calibration \cite{bib:antares_tcalib} of the PMTs was performed at the integration laboratories before the deployment of each line. A laser device sending light to the PMTs through an optical fiber was used. 
After correcting for the propagation delay within the line, the time offsets between each ARS and an ARS chosen as a reference were calculated. These  offsets defined a first set of time calibration parameters, usually referred to as the \texttt{ARS$\_$T0} or the {\it intra line} calibration parameters. These values were stored in the ANTARES database and used for the prompt data analysis after deployment.

The value of \texttt{ARS$\_$T0} measured in the laboratory may change after the line deployment due to, e.g., variations in the environmental conditions.  Therefore, a system allowing for the {\it in situ} time calibration of the detector  and the monitoring of the calibration constants is needed. The optical beacon system can be used to perform the relative time calibration of the optical sensors. There are two kinds of optical beacons: the LED beacons and the laser beacons. The LED optical beacons consist of 36 individual blue LED light sources arranged in groups of six on electronic boards placed side by side, forming a hexagonal prism, enclosed in a glass container.  Four LED optical beacons are positioned regularly (in the 2nd, 9th, 15th and 21st storeys from the bottom) on each detector line. The non-consecutive LED optical beacon pairs within a line are fired simultaneously at maximum intensity during a calibration run. In total, 24 runs of 5 minutes duration are taken once per month. The runs are analysed later to check and update (if needed) the \texttt{ARS$\_$T0} parameters. 

In order to measure the relative offsets between the detector lines (\textit{inter line} calibration), one laser beacon is placed at the bottom of a central line. This device consists of a Nd-YAG laser, which emits green light in high-intensity short pulses (FWHM $\sim$1 ns) illuminating up to the 10th floor of every detector line. Currently, one laser beacon run of 5-10 minutes duration is taken every month. However, it is important to complement this system with the method proposed here, as mentioned in Sec. \ref{sec:inter_line}.


\section{Track reconstruction}
\label{sec:track_reco}
\begin{figure}
 \begin{center}
\includegraphics[width=0.5\linewidth]{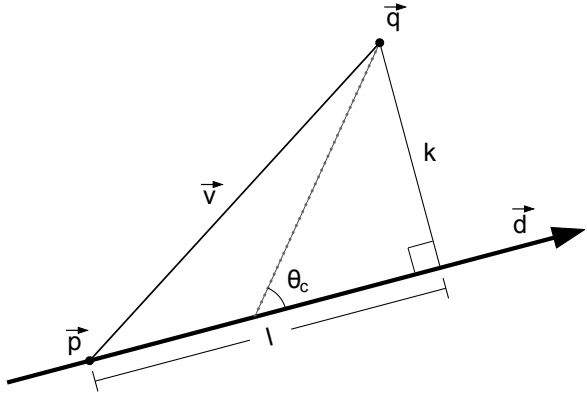}
\caption{ A muon passing near the detector along a direction $\vec{d}$ and a position $\vec{p}$ at a given time $t_0$ induces in the medium the production 
of Cherenkov photons. These photons are emitted at an angle $\theta_{c}$ with respect to the muon trajectory, $\vec{d}$. They are detected by the PMT located at position $\vec{q}$ after travelling the light-grey line path. $l$ and $k$ indicate the components of the vector $\vec{v} = \vec{q}-\vec{p}$ in the parallel and perpendicular directions with respect to $\vec{q}$, respectively.}
\label{fig:residuals_geo}
 \end{center}
\end{figure}

The muon track reconstruction algorithm \cite{bib:aart_reco} used by ANTARES both for physics analyses and for timing calibration 
consists of several fitting procedures that are performed consecutively.
In the final step, estimates of all the parameters needed to describe the event trajectory 
are obtained by maximising the likelihood function 
\begin{equation}
\mathcal{L}(\overrightarrow{d},\overrightarrow{p}) = \prod P (t_{i} | t^{th}_{i},\overrightarrow{d},\overrightarrow{p}).
\label{eq:likelihood}
\end{equation}
$\mathcal{L}$ is the product of the likelihood, $P$, of the individual hits accounting for the 
probability of the time of the hits, 
where $\vec{p}$ indicates a point in the muon track direction $\vec{d}$ at a time $t_0$ and $t_{i}$ indicates the measured hit time. $t^{th}$ is the expected hit time given by

\begin{equation}
t^{th}_{i} = t_{0} + \frac{1}{c}(l - \frac{k}{\tan\theta_{c}}) + \frac{1}{v_{g}}(\frac{k}{\sin\theta_{c}}) ,
\label{eq:residuals}
\end{equation}

\noindent derived from a given set of parameters 
(see Fig. \ref{fig:residuals_geo}), namely the position, $\vec{q}$, of the PMT which has detected the Cherenkov photon; the components of the vector $\vec{v} = \vec{q}-\vec{p}\,$ in the parallel, $l$, and perpendicular, $k$, directions to the muon track; the group velocity of light, $v_g$; the angle of the Cherenkov light, $\theta_c$; and the speed of light in vacuum, $c$. 

The likelihood in Eq. \ref{eq:likelihood} can be expressed in terms of the probability density 
of the time residuals $r_i = t_{i} - t_{i}^{th}$. 
The probability density function used for the likelihood fit is obtained from simulations, 
and takes into account the contribution from hits arriving late due to Cherenkov emission by 
secondary particles and/or light scattering, and 
the effect of the transit time spread of the PMT. The probability of a hit being 
due to background is also accounted for. 

The quality of the track fit is quantified by the parameter

\begin{equation}
\Lambda = \frac{\log(\mathcal{L})}{N_{{DOF}}} + 0.1  (N_{{comp}}-1),
\label{eq_lambda}
\end{equation}

\noindent which incorporates the maximum likelihood value $\mathcal{L}$, the number of degrees of freedom in 
the fit $N_{{DOF}} = N_{{hits}} - 5$, equal to the number of hits used minus the number 
of free parameters. $N_{{comp}}$  is the number of compatible solutions found by the maximization 
algorithm, which in general takes large values for well-reconstructed tracks and small ones for
badly reconstructed events. 
The $\Lambda$ parameter is a useful tool to reject badly reconstructed events, mainly  
down going atmospheric muons misreconstructed as up going.
Moreover, the uncertainty on the reconstructed track direction is used for further event selection. 
Assuming that the likelihood function near the fitted maximum follows a multivariate Gaussian distribution, 
the error on the zenith ($\beta_{\theta}$) and azimuth ($\beta_\phi$) angles can be derived from the likelihood fit covariance error matrix. 
From these errors, the parameter
\begin{equation}
\beta = \sqrt{ \sin^{2}(\theta_{{rec}})\beta^{2}_{\phi} + \beta^{2}_{\theta} },
\end{equation}
referred to as the angular error estimate, is obtained, where $\theta_{rec}$ is the reconstructed zenith angle. A cut on the angular error estimate is 
highly efficient for signal events and removes a large fraction of misreconstructed 
atmospheric muons \cite{bib:aart_reco}.

%

\section{Time calibration with muon tracks: method}
\label{sec:method_desc}
\begin{figure}
 \begin{center}
\includegraphics[width=1.0\linewidth]{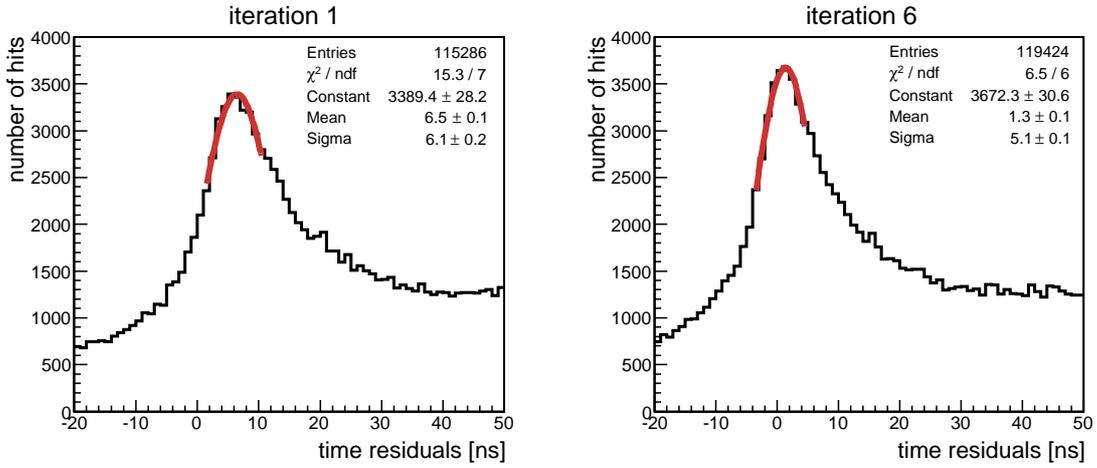}
\caption{Distributions of the time residuals for hits detected with line 8 after a first iteration 
of the calibration method (left) and after applying six iterations on the same data sample (right). A Gaussian function is fitted to each distribution around the position of the peak. The 
RMS of the distribution decreases, and the number of entries increases, with increasing iterations, since the 
timing corrections applied at each new iteration have a positive effect on the event reconstruction.} 
\label{fig:time_res_gauss}
 \end{center}
\end{figure}

\begin{figure}
 \begin{center}
\includegraphics[width=0.6\linewidth]{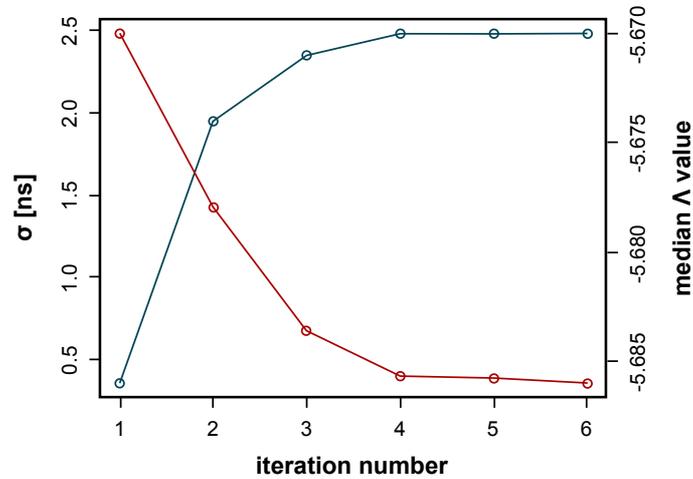}
\caption{Width of the inter line distribution of time corrections as a function of the number of 
iterations of the procedure (red line and left-hand side scale). The median value of the fit quality parameter distribution of the 
events increases with the number of iterations, as expected (blue line and right-side scale).} 
\label{fig:width_reco_iters}
 \end{center}
\end{figure}

The reconstructed trajectories of the down going muons produced in cosmic-ray interactions in the atmosphere are used for the time calibration of the ANTARES detector. These are collected at a rate of 
$\sim$5 Hz. 
The method  is an iterative procedure that uses  the time residual distributions obtained 
for a subset of hits which are not used in the track fit. The complete procedure consists of the following steps:

\begin{enumerate}

\item A subset of hits (the {\it probe hits}), which were registered  by the ARS or were within the line under investigation,  is selected. These hits are  not used in the track fit. 
\item The muon trajectory is reconstructed using only the remaining hits (the {\it reco hits}).
\item Time residuals for the probe hits are calculated with respect to the fitted track. 
\item A Gaussian function is fitted to the peak of the probe hit time residuals distribution (an example is given in Fig. \ref{fig:time_res_gauss}), the mean value of which is taken as the time correction 
to be applied to the hit times for a new iteration of the method. 

\end{enumerate}
The exclusion of the probe hits guarantees that the resulting time residual distributions are not biased, since the muon track has not been fitted to minimise them. 
If  the subset of probe hits includes only those collected on one single ARS, the so-called intra line calibration is performed. If it includes those collected on one detection line,  the inter line calibration is performed.
This procedure is repeated until the time corrections are small ($\sim$ 0.5 ns). 
The reconstructed muon tracks used for the time residuals are affected by the time corrections applied; at each new iteration, the number of high quality reconstructed events increases with the number of iterations until the method converges to an optimal solution. Fig. \ref{fig:width_reco_iters} shows how fast this solution convertes to its noise limit of 0.5 ns when applying the method to determine the time offsets between different detection lines. In the same figure it is also shown how the median value of the fit quality distribution increases with the number of 
iterations of the method.

\section{Results on the inter line timing}
\label{sec:inter_line}
The  inter line calibration corrects the time offsets with respect to a common reference for the full detector. Since at the integration laboratories the time calibration is only performed at the line level, this is the most important calibration performed off-shore. Indeed, the first hint of the existence of time offsets between the  lines of the ANTARES detector was found when studying the distributions of the quality reconstruction parameter $\Lambda$.  A large discrepancy between data and Monte Carlo simulations\footnote{In the ANTARES collaboration, the generation of atmospheric 
neutrinos is performed with the GENHEN package \cite{bib:antares_genhen} using the Bartol model \cite{bib:antares_bartol}, while the atmospheric muons are produced with MUPAGE \cite{bib:antares_mupage}.} 
was found for tracks crossing the detector diagonally and inducing hits on several lines, whereas nearly vertical tracks, the reconstruction of which is single-line dominated, showed better agreement. 

In order to determine the time offsets, the method described in the previous section is here applied to data recorded with 12 detection lines in a few days of data-taking from March 2010. Other periods have been studied (November 2010, April 2011, March 2012) with no significant differences in the time offsets. Good quality events  (only tracks fitted with $\Lambda > -6.0$) were used.
Fig. \ref{fig:time_res_gauss} shows the time residuals distribution, after one (left) and after six (right) iterations of the method, for hits detected on line 8, which is 
one of the central lines of the ANTARES detector. The  range of the Gaussian fit applied to these histograms was chosen in order to avoid the contribution from scattered photons and to match the most Gaussian-like region of the residuals distribution.

The inter line constants determined for the 12 detection lines are summarised in Table 
\ref{tab:line_offsets}. 
These are the values normally used in ANTARES analyses.


\begin{table}
\begin{center}
\begin{tabular}{c|cccccccccccc} 

Line & L1 & L2 & L3 & L4 & L5 & L6  &  L7 & L8 & L9 &L10 & L11 & L12\\
\hline
Offset (ns) & 1.3 & -3.9 & -0.5 & -2.1 & -3.4 & -1.3 & 0.6 & 4.9 & 0.3 & -0.5 & 4.0 & 2.2\\
\end{tabular}
\caption{Inter line offsets measured with the track residual method. These values are cu\-rrent\-ly used for data 
processing.}
\label{tab:line_offsets}
\end{center}
\end{table}

\subsection{Impact on the reconstruction and on the angular resolution.}
\label{sec:Impact-reco}
\begin{figure}
 \begin{center}
\includegraphics[width=1\linewidth]{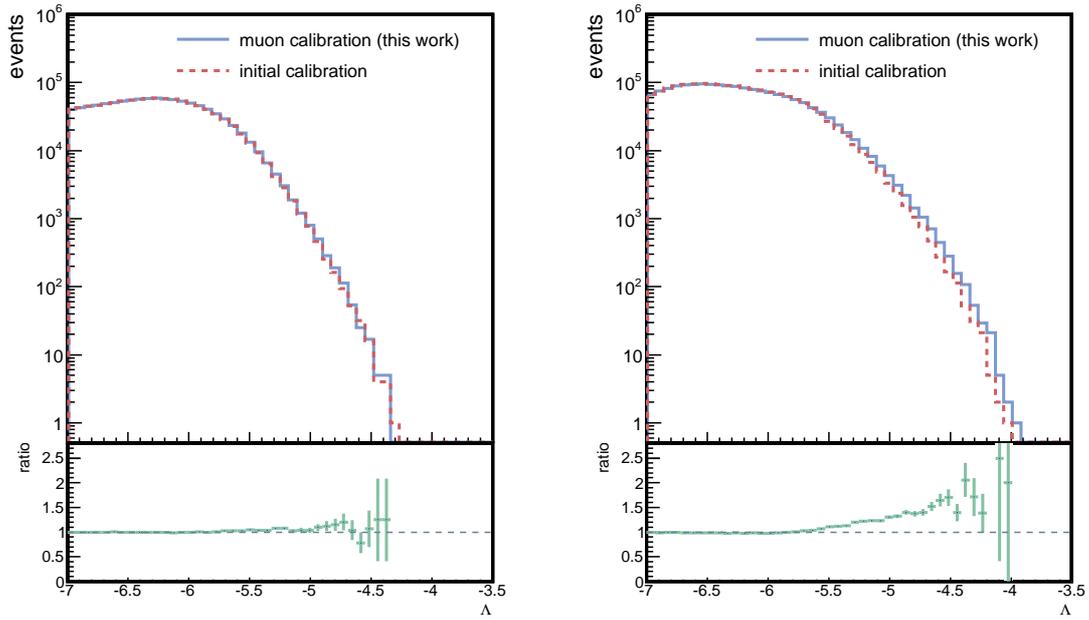}
\caption{Distributions of the reconstruction quality parameter before and after correcting 
the inter line offsets for (left) vertically down going tracks, $\cos\theta <-0.9$, 
and (right) inclined tracks, $\cos\theta >-0.8$. The lower pads show the ratio between 
the number of events reconstructed after and before applying the inter line timing correction. For inclined tracks, the number of very well reconstructed tracks improves a factor 1.2-2.0 for $\Lambda >$ -5.3.}
\label{fig:fit_quality_diag_and_vert}
 \end{center}
\end{figure}

After the application of the corrected inter line timing, an improvement in the quality of the reconstruction was observed with an increase of the number of events with high $\Lambda$ values. 
The improvement was particularly important (see Fig. \ref{fig:fit_quality_diag_and_vert}) 
for events reconstructed on several lines,  
while for vertically down going events, which are
mainly single-line reconstructed, the improvement was smaller, as expected.
The agreement of the   distributions of the $\Lambda$ parameter for data and simulated events improved after the timing calibration correction by  $\sim$ 15$\%$ for well-reconstructed tracks

\begin{figure}
 \begin{center}
\includegraphics[width=1\linewidth]{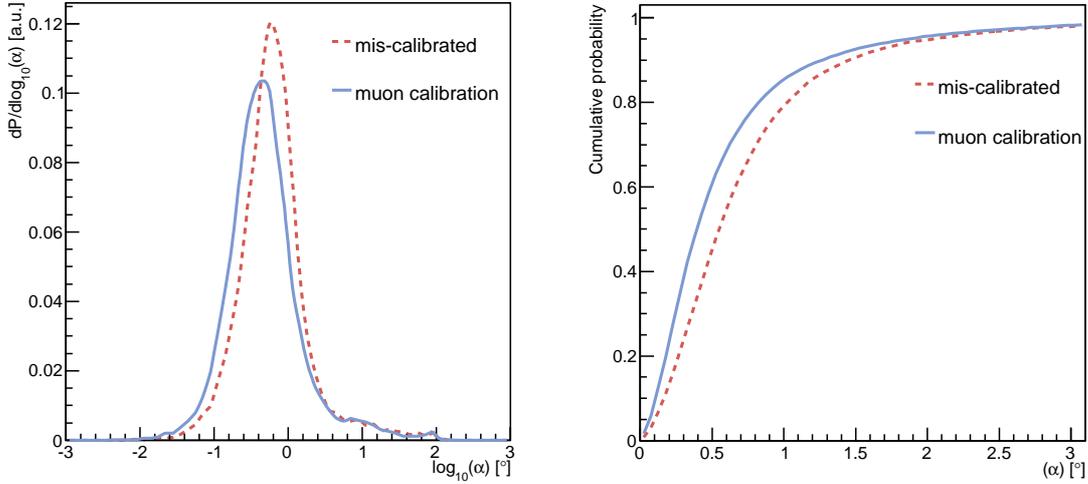}
\caption{Left: angular error for the simulated tracks with mis-calibrated timing (dashed line) and standard simulation 
(solid line). Right: the corresponding cumulative distributions.}
\label{fig:angular_error_miscal_lines_mc}
 \end{center}
 \vspace{1cm}
\end{figure}

In order to study the effect of the inter line time shifts on the detector resolution, offsets of the same amounts as the values listed in Table \ref{tab:line_offsets} were added to the hit times of simulated events and the reconstructed tracks compared to the results of reconstruction from the usual simulation.
Fig. \ref{fig:angular_error_miscal_lines_mc} shows the distribution of the angular difference, $\Psi$, between the reconstructed track direction and the generated one for 
the default Monte Carlo simulation and for events reconstructed using mis-calibrated lines. 
Adding the offsets results in a resolution degraded by about $40\%$ degraded resolution for the selected events 
($\Lambda > -5.4 \; $, $\; \cos\theta_{rec} > 0.0 \;$ and $ \; \beta < 1^{\circ}$). 
The median value of the distribution worsens from 0.40$^{\circ}$ to 
0.55$^{\circ}$ for the studied sample.

\newpage
\subsection{Cross-check with simulations}

\begin{figure}[!h]
 \begin{center}
\includegraphics[width=0.6\linewidth]{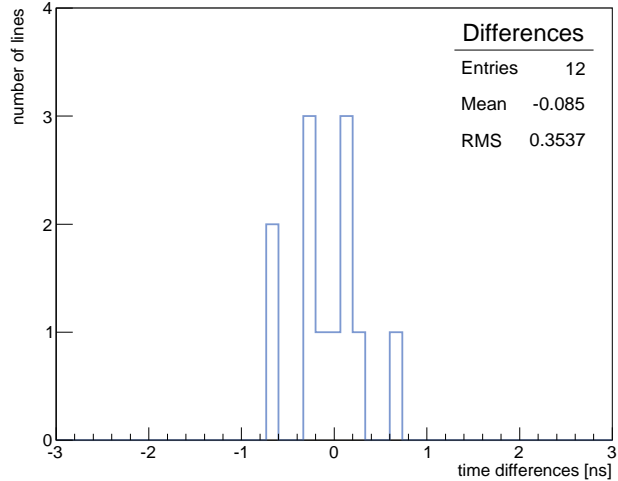}
\caption{Distribution of the differences between the inter line offsets determined by 
applying the procedure on Monte Carlo simulations and the values introduced in order to distort the timing of the lines.}
\label{fig:mc_offsets}
 \end{center}
\end{figure}

The precision of the muon time residuals method can be tested using simulations by adding inter line  time offsets in the event reconstruction chain. The same time offsets as in the values summarised in 
Table \ref{tab:line_offsets} were introduced to distort the timing measured by the lines. 
The resulting simulated reconstructed tracks had the muon time residuals method applied to determine 
the  inter line timing offsets.
Fig. \ref{fig:mc_offsets} shows the measured time offsets compared 
 with the offsets artificially added to mis-calibrate the lines. 
The RMS of the distribution is 0.35 ns, which is well within the required precision.

\subsection{Cross-check with the optical beacon system}

The laser beacon can also be used as well to determine the time offsets between the detector lines. 
This system provides an independent calibration method to cross-check 
the results obtained using the information provided by the muon track residuals. For this purpose, a calibration run is taken every month in the ANTARES detector. 
The laser, at the bottom of line 8, illuminates the detector for about 10 minutes.
The data are then analysed following a similar procedure to the one
applied for the \texttt{ARS$\_$T0} in-situ calibration \cite{bib:antares_tcalib}.
The method is based on the study of the time differences between the emission
of the light pulse and the time when  light is detected by the OMs. 
Time residuals are calculated by correcting for the required photon propagation time. 
Gaussian functions convolved with an exponential are fitted to the time residual distributions. The obtained fit peak values 
are then plotted as a function of the distance between the OM and the laser beacon 
position. Only those OMs which are illuminated by the laser at the single photoelectron 
level are used in the calculation, because for them a constant relation between 
residual peak positions and distance to the light source is expected. 

\begin{figure}
 \begin{center}
\includegraphics[width=0.6\linewidth]{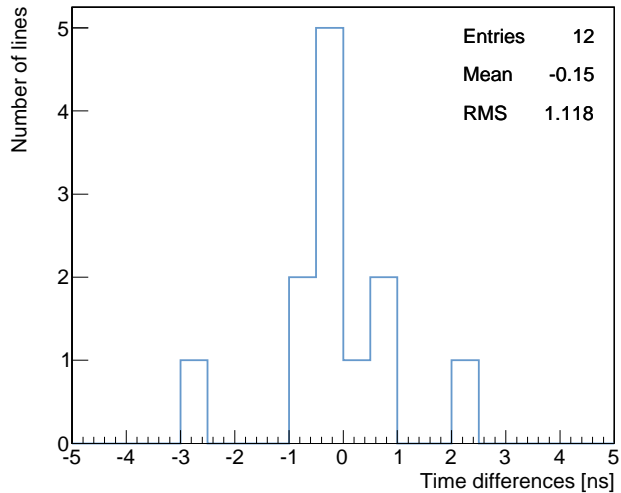}
\caption{Time difference between the inter line
offsets measured using the reconstructed muon tracks and the values obtained 
using the laser beacon system.}
\label{fig:muon_ob_comparison}
 \end{center}
\end{figure}

In Figure \ref{fig:muon_ob_comparison} the inter line offsets measured with the laser optical beacon are compared to the results obtained using the track residuals method. 
They agree to within 1 ns, except for line 1 and line 8, which both show a discrepancy larger than 2 ns. 
The discrepancy for line 1 may be due to the large distance from  the laser optical beacon (line 1 is the farthest), and for line 8,  to a shadowing effect that prevents light from  directly reaching the uppermost OMs.

\begin{figure}
 \begin{center}
\includegraphics[width=1.0\linewidth]{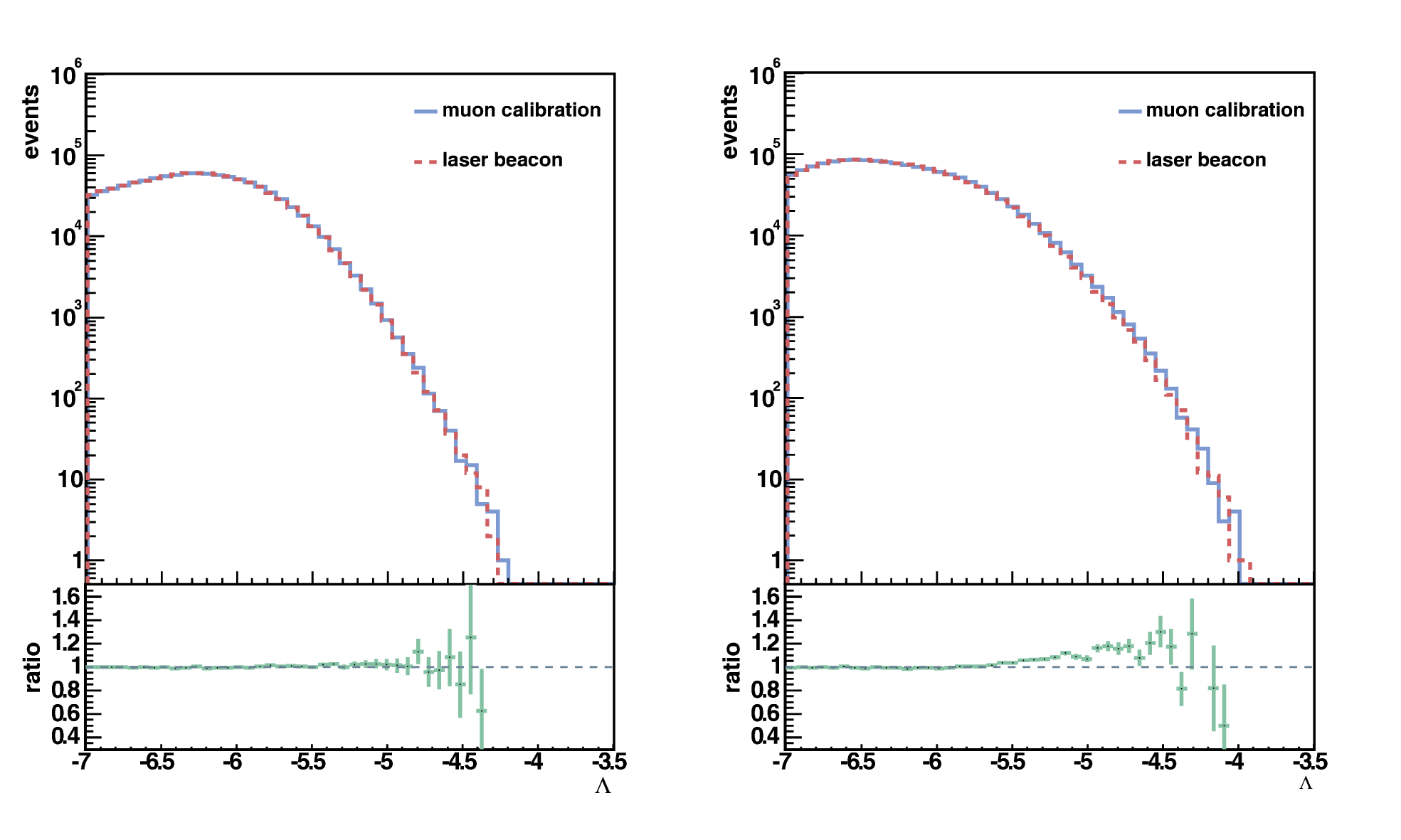}
\caption{Distribution of the quality of the reconstruction
parameter for events that have been reconstructed with left: $\cos(\theta)<-0.9$ and right: $\cos(\theta)>-0.8$,
using the inter line offset corrections provided by the track residuals method (solid line) and
those obtained using the laser beacon data (dashed line). In each case, the bottom plot shows the ratio between the
number of reconstructed events for each case (muons/laser). For cos$\theta >$ --0.8, a $>$10\% improvement can be seen from $\Lambda >$ -5.3. }
\label{fig:lambda_muon_vs_laser}
 \end{center}
\end{figure}

It was found by examining the reconstructed events that the track residuals method is more efficient than the laser optical beacon method, producing a slightly higher number of well reconstructed tracks
(see Figure \ref{fig:lambda_muon_vs_laser}).

\section{Results on the intra line timing}
\label{sec:intra_line}
\begin{figure}
 \begin{center}
\includegraphics[width=0.6\linewidth]{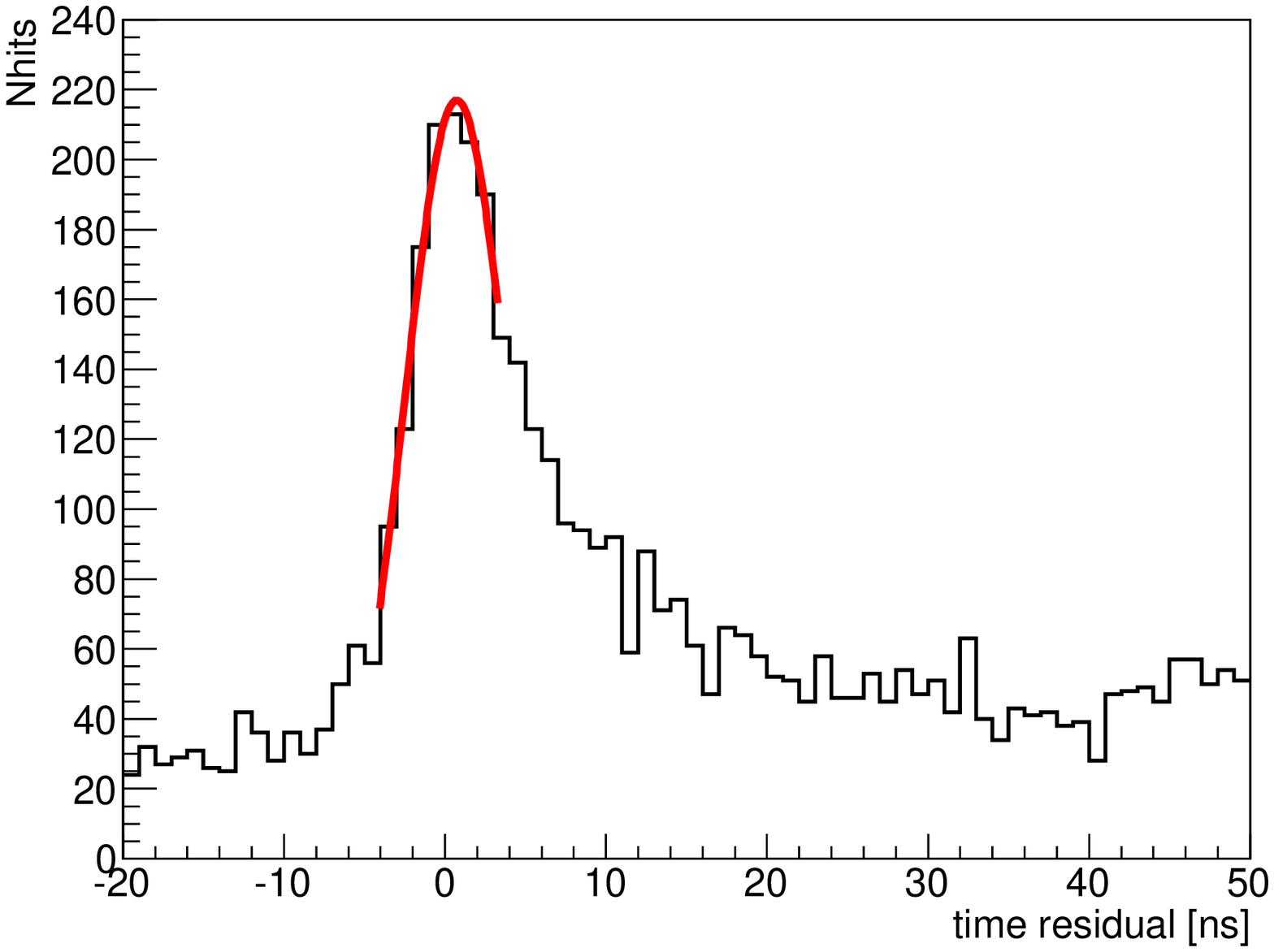}
\caption{Distribution of the time residuals for ARS number 3 in line 4, floor 10. 
A gaussian function is fitted around the peak of the distribution, the mean of which is interpreted 
as the ARS time offset.}
\label{fig:example_ars_tres_dist}
 \end{center}
\end{figure}

The intra line calibration with atmospheric muons is performed by producing  
time residual distributions for every ARS in the detector. In order to 
reduce the amount of required data, the first step in the scheme 
explained in section \ref{sec:method_desc} is modified to provide probe-hit time residuals 
for those ARSs that have registered at least one hit from the current event. 
Again, the resulting distributions are fit with a Gaussian function, the mean of which is taken 
as the time offset of the ARS under study. As an example, Figure \ref{fig:example_ars_tres_dist} shows the  fitted
probe-hit residuals distribution for one particular ARS in line 4 and floor 10.

\begin{figure}
\begin{center}
\includegraphics[width=0.8\linewidth]{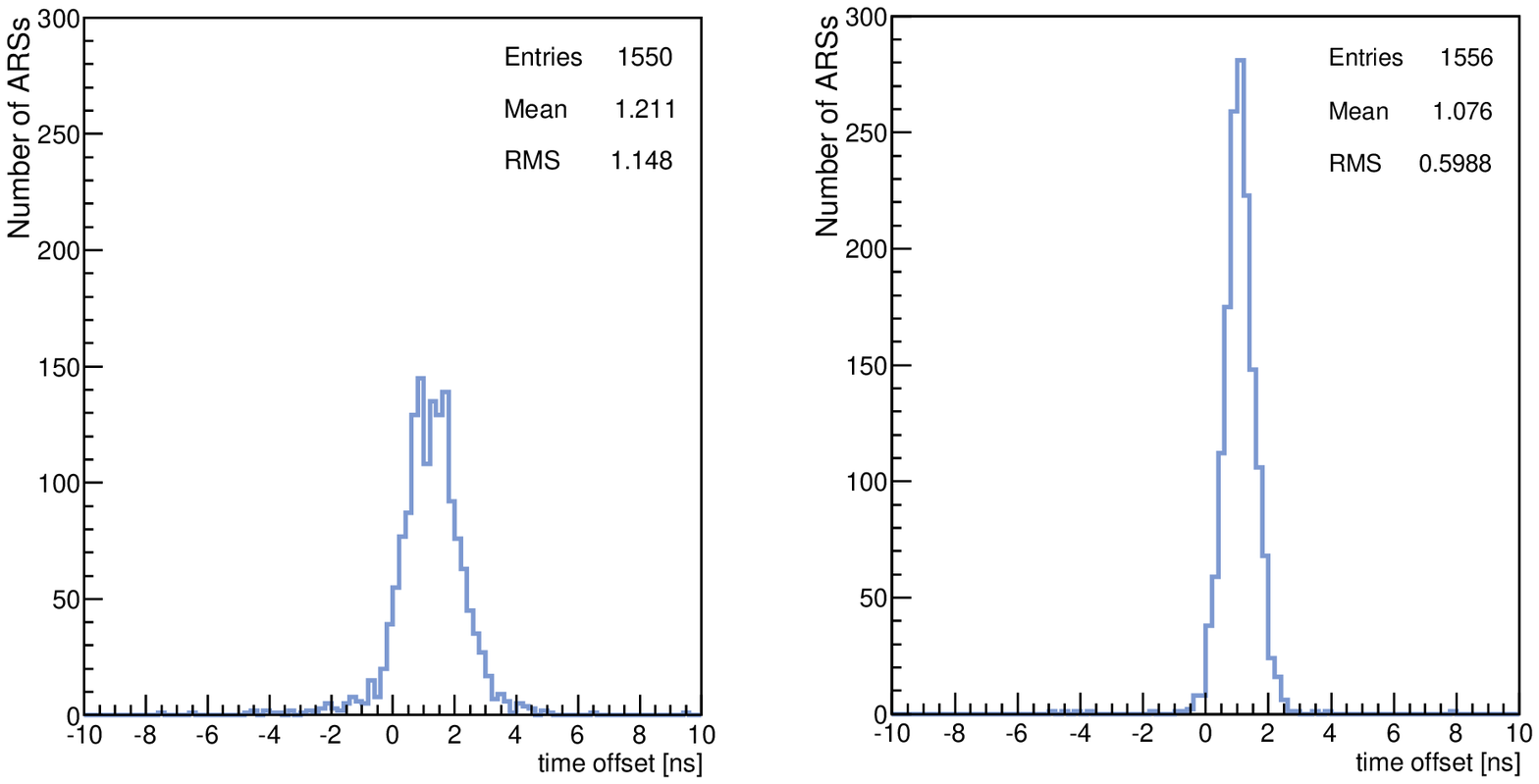}
\caption{Distribution of the ARS timing corrections after one iteration (left) and after five 
iterations (right) of the muon time residuals method. The increase in the number of entries is due to a few ARS which are outside the time range in the left figure.}
\label{fig:ars_toff_corrections}
\end{center}
\end{figure}

Since the \texttt{ARS$\_$T0} constants provided by the LED beacon calibration system are taken into account 
in the muon track reconstruction, the procedure discussed in this paper provides corrections to the intra line 
calibration parameters obtained using the beacons.
In Figure \ref{fig:ars_toff_corrections} the distribution of these
corrections is shown including all ARSs with sufficient statistics after one iteration (left) 
and after five iterations (right) of the method, when the procedure has converged to a solution.

\subsection{Effect on the reconstruction}
\begin{figure}
\begin{center}
\includegraphics[width=0.6\linewidth]{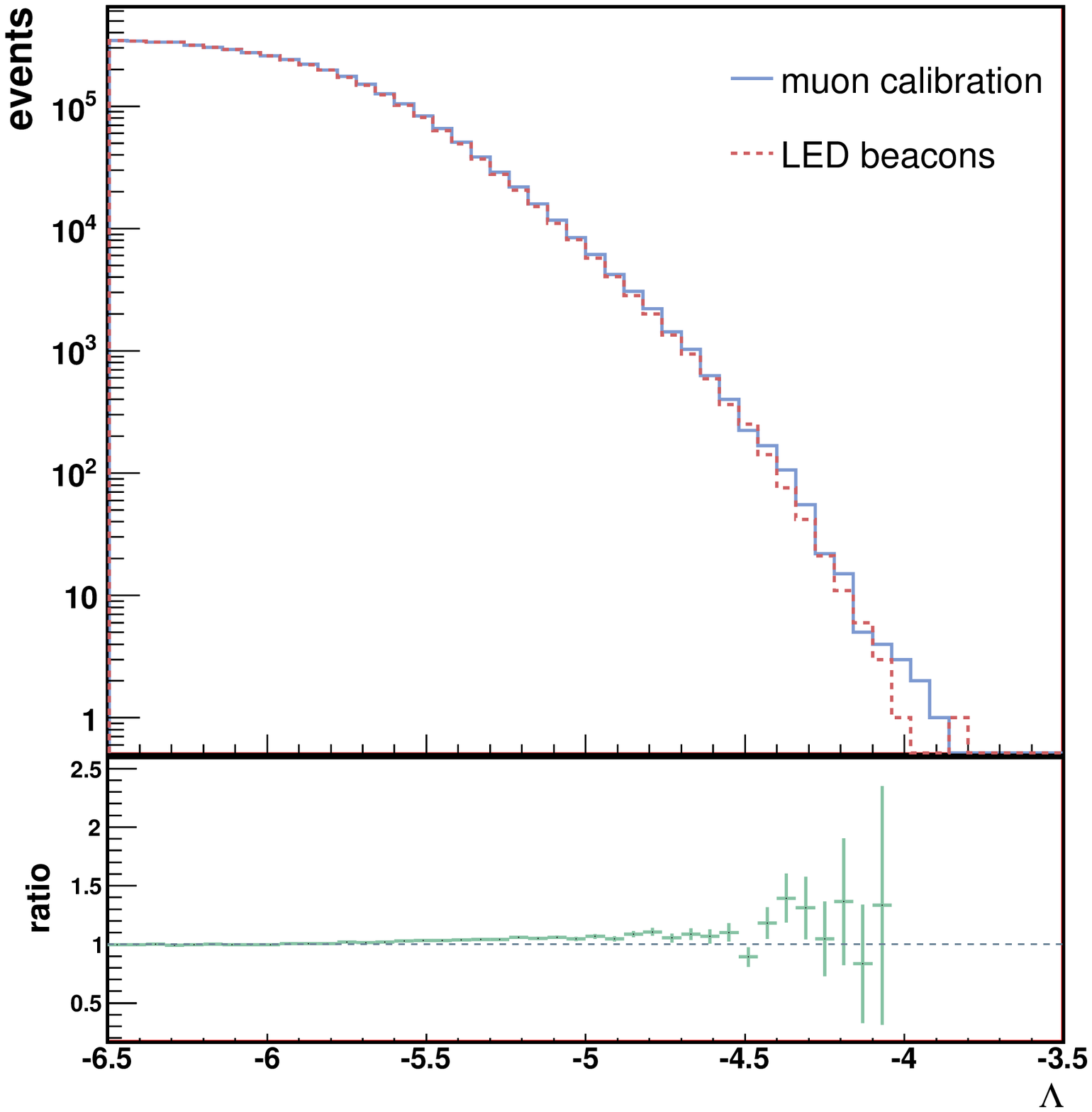}
\caption{Distributions of the quality of the reconstruction parameter for events reconstructed using the LED 
beacon intra line calibration constants (dashed line) and reconstructed by taking into account the corrections 
determined by applying the muon time residuals method (solid line). Shown in the lower pad is the ratio between 
the events reconstructed by applying the muon time residuals method and the events reconstructed using the 
LED beacon calibration constants only.}
\label{fig:track_params_intra_line}
\end{center}
\end{figure}

In order to determine the effect on the reconstruction of the measured ARS time corrections, 
a small sample of data runs 
has been analysed. 
In Fig. \ref{fig:track_params_intra_line} a comparison 
of the reconstruction quality parameter for events reconstructed considering the LED beacon calibration 
(dashed line) and reconstructed applying the time corrections obtained using the method presented in this paper is shown. 
A small increase in the number of high quality events is observed, similar to the findings from simulations in Section \ref{sec:Impact-reco}, after correcting for artificially introduced time offsets. This validates the intra-line-timing calibration produced with the muon time residuals method.
Note also that, while the LED beacons typically calibrate about one thousand ARSs, the muon tracks method allows the timing corrections of almost all of the active channels (nearly 1600 in this study) to be obtained.

\section{Conclusions}
\label{sec:conclusions}
A  method for the time calibration of the ANTARES detector based on the reconstruction of atmospheric muons has been presented. It has been used to measure the 
relative time offsets between the different detection lines, already measured in the laboratory, with a precision of 1 ns. A calibrated  inter line  timing 
results in a factor of 1.2-2.0 improvement in the number of high-quality events reconstructed 
in the zenith angle range $-0.8<\cos(\theta)<0$, i.e.  events traversing the detector diagonally.
It has also been shown that not accounting for the inter line time corrections would result in 
a degradation of the expected detector resolution by approximately 40\%.
A comparison with a timing calibration results obtained with the laser beacon system shows that the two methods are compatible to within 1 ns. However, the 
muon time residuals procedure is more efficient  for the improvement of the quality of reconstructed events and allows the evaluation of the  offsets for two  lines that  cannot be considered correctly 
with the laser optical beacon.
The muon time residuals method has also been applied to measure corrections for the  intra line timing 
constants first determined using the system of LED optical beacons. The results show a slight 
improvement in the quality of the reconstruction. The calibration explained here is applied in ANTARES analyses.

\section*{Acknowledgments}

This article is dedicated to Juan Pablo G\'omez-Gonz\'alez, whose contribution has been instrumental to this paper.  Juan Pablo was by mistake the target of a vile assault and battery whose physical consequences were extremely serious. The determination with which Juan Pablo has faced his fate and overcome the moral and physical damage that he has suffered is profoundly admirable. The ANTARES Collaboration wish to express their solidarity with Juan Pablo, and their great sympathy and regard for him.

The authors acknowledge the financial support of the funding agencies:
Centre National de la Recherche Scientifique (CNRS), Commissariat \`a
l'\'ener\-gie atomique et aux \'energies alternatives (CEA),
Commission Europ\'eenne (FEDER fund and Marie Curie Program), R\'egion
\^Ile-de-France (DIM-ACAV) R\'egion Alsace (contrat CPER), R\'egion
Provence-Alpes-C\^ote d'Azur, D\'e\-par\-tement du Var and Ville de La
Seyne-sur-Mer, France; Bundesministerium f\"ur Bildung und Forschung
(BMBF), Germany; Istituto Nazionale di Fisica Nucleare (INFN), Italy;
Stichting voor Fundamenteel Onderzoek der Materie (FOM), Nederlandse
organisatie voor Wetenschappelijk Onderzoek (NWO), the Netherlands;
Council of the President of the Russian Federation for young
scientists and leading scientific schools supporting grants, Russia;
National Authority for Scientific Research (ANCS), Romania; 
Mi\-nis\-te\-rio de Econom\'{\i}a y Competitividad (MINECO), Prometeo 
and Grisol\'{\i}a programs of Generalitat Valenciana and MultiDark, 
Spain; Agence de  l'Oriental and CNRST, Morocco. We also acknowledge 
the technical support of Ifremer, AIM and Foselev Marine for the sea 
operation and the CC-IN2P3 for the computing facilities.








\end{document}